\documentclass[acmsmall]{acmart}

\usepackage{calc}
\usepackage{graphicx} 
\usepackage{booktabs}
\usepackage{hyperref}
\usepackage{listings}
\usepackage[table]{xcolor}
\usepackage{float}
\usepackage{caption}
\usepackage{newfloat}  
\usepackage[most]{tcolorbox}
\usepackage{enumitem}
\usepackage{multirow}
\usepackage{pifont} 
\usepackage{balance}
\usepackage[normalem]{ulem}
\usepackage{xurl} 
\usepackage{siunitx}

\DeclareFloatingEnvironment[fileext=frm,placement=h,name=Figure]{codefigure}

\definecolor{lightyellow}{rgb}{1.0, 1.0, 0.6}

\definecolor{lightgray}{RGB}{211, 211, 211}

\lstdefinelanguage{JavaScript}{
  keywords={break, case, catch, continue, debugger, default, delete, do, else, finally, for, function, if, in, instanceof, new, return, switch, this, throw, try, typeof, var, void, while, with, let, const, await, async},
  numbers=left,
  numberstyle=\tiny\color{gray},
  stepnumber=1,
  numbersep=5pt,
  showstringspaces=false,
  breaklines=true,
  frame=lines,
  morecomment=[l]{//},
  morecomment=[s]{/*}{*/},
  morestring=[b]",
  morestring=[b]',
  sensitive=true
}

\lstdefinelanguage{Python}{
  keywords={and, as, assert, break, class, continue, def, del, elif, else,
            except, False, finally, for, from, global, if, import, in, is,
            lambda, None, nonlocal, not, or, pass, raise, return, True, try,
            while, with, yield},
  numbers=left,
  numberstyle=\tiny\color{gray},
  stepnumber=1,
  numbersep=5pt,
  showstringspaces=false,
  breaklines=true,
  frame=lines,
  morestring=[b]',
  morestring=[b]",
  morecomment=[s]{"""}{"""},  
}

\lstdefinelanguage{TypeScript}{
  language=JavaScript,
  morekeywords={interface, type, implements, private, public, protected, readonly, string, number, boolean, any, never, unknown}
}

\lstdefinelanguage{json}{
    morestring=[b]",
    showstringspaces=false,
    backgroundcolor=\color{white},
    stringstyle=\color{black},
}

\lstdefinelanguage{text}{
    morestring=[b]",
    showstringspaces=false,
    breaklines=true,
    backgroundcolor=\color{white},
    stringstyle=\color{black},
}

\tcbset{
  takeaway/.style={
    enhanced,
    colback=gray!10,       
    colframe=gray,        
    boxrule=0.3pt,         
    arc=3pt,               
    boxsep=1pt,            
    left=2pt,
    right=2pt,
    top=2pt,
    bottom=2pt,
    width=\columnwidth,    
  }
}

\definecolor{zwbcolor}{rgb}{0,0.5,0.5}

\definecolor{clcolor}{rgb}{0.5,0.7,0.9}

\definecolor{pxcolor}{rgb}{0.8,0.6,0.9}

\definecolor{rthcolour}{rgb}{0.2,0.2,0.8}

\definecolor{revblue}{rgb}{0,0,1}
\newcommand{\rev}[1]{\textcolor{revblue}{#1}}
\renewcommand{\rev}[1]{#1}
\newcommand{\revdel}[1]{\textcolor{red}{\sout{#1}}}
\renewcommand{\revdel}[1]{}

\newcommand{\toolname}[0]{\textsc{DocPrism}}

\setcopyright{cc}
\setcctype{by}
\acmDOI{10.1145/3832248}
\acmYear{2026}
\acmJournal{PACMSE}
\acmVolume{3}
\acmNumber{ISSTA}
\acmArticle{ISSTA157}
\acmMonth{10}
\acmSubmissionID{issta26main-p1715-p}
\received{2026-01-29}
\received[accepted]{2026-06-25}
\begin{document}






\title{DocPrism: Multi-lingual Detection of Incorrectness Inconsistencies between Code and Documentation}

\author{Xiaomeng Xu}
\orcid{0009-0008-4842-5426}
\affiliation{%
  \institution{University of British Columbia}
  \city{Vancouver}
  \country{Canada}
}
\email{xmxu@cs.ubc.ca}

\author{Zahin Wahab}
\orcid{0009-0008-7273-4165}
\affiliation{%
  \institution{University of British Columbia}
  \city{Vancouver}
  \country{Canada}
}
\email{zwb@cs.ubc.ca}

\author{Reid Holmes}
\orcid{0000-0003-4213-494X}
\affiliation{%
  \institution{University of British Columbia}
  \city{Vancouver}
  \country{Canada}
}
\email{rtholmes@cs.ubc.ca}

\author{Caroline Lemieux}
\orcid{0000-0002-9610-8520}
\affiliation{%
  \institution{University of British Columbia}
  \city{Vancouver}
  \country{Canada}
}
\email{clemieux@cs.ubc.ca}


\markboth{Journal of xxxxx ,~Vol.~xxx, No.~xxx, xxxxxxx}%
{Shell \MakeLowercase{\textit{et al.}}: A Sample Article Using IEEEtran.cls for IEEE Journals}


\begin{abstract}
  Code-documentation inconsistencies are common and undesirable: they  can lead to developer misunderstandings and software defects. This paper introduces \toolname{}, a lightweight multi-language, code-documentation inconsistency detection tool. \toolname{} uses a standard large language model (LLM) to analyze and explain inconsistencies, and focuses on outputting \emph{incorrectness} inconsistencies. Plain use of LLMs for this task yields unacceptably high inconsistency flag rates---i.e., over 90\% of functions are flagged as inconsistent with their documentation. One  substantial reason is that LLMs identify natural gaps between high-level documentation and code as \emph{incompleteness}  inconsistencies. We introduce and apply the \emph{Local Categorization, External Filtering} (LCEF) methodology: 
  LCEF uses an LLM's local completion skills, rather than its long-term reasoning skills, to focus on reporting \emph{incorrectness} inconsistencies.
  In our ablation study, LCEF reduces \toolname{}'s inconsistency flag rate from 98\% to 14\%, 
and increases F1 score from 0.22 to 0.77, compared to standard prompting techniques. 
  On a broad evaluation across 
  Python, TypeScript, C++, and Java, \toolname{} maintains a low flag rate of 17\%, and achieves a precision of 0.63 without performing any 
  fine-tuning. We also establish a conservative lower bound across four programming languages, showing that inconsistency errors are present in $11\%$ of code-documentation pairs. \rev{In addition, \toolname{} achieves precision comparable to the state-of-the-art on an established synthetic dataset, but substantially outperforms it on our real-world Java dataset in precision (\toolname{}: 0.47–0.67 vs. SOTA: 0.05–0.14).} 
\end{abstract}

\ccsdesc[500]{Software and its engineering~Documentation}
\ccsdesc[300]{Software and its engineering~Software maintenance tools}

\keywords{code-documentation inconsistency, documentation, code comments, LLM}

\maketitle


\section{Introduction} 

\label{sec:intro}
Method-level documentation acts as a specification describing the behaviours a method provides. 
Documentation serves as an abstraction developers use to understand how to correctly use a method without having to read through its source code.
Unfortunately, long experience has taught developers not to trust documentation, mainly due to inconsistencies between what the documentation says and what the code actually does. 
The goal of this paper is to \emph{detect} and \emph{explain} such code-documentation inconsistencies.

This work focuses on surfacing \emph{incorrectness} inconsistencies between code and documentation. First, cases where there is a logical conflict between code and its documentation: we call this a \emph{direct mismatch}. Second, cases where a behaviour is documented but not implemented: we call this an \emph{over-promise}. Both of these are true incorrectness inconsistencies: they imply that either the code or its documentation is erroneous. 

Identifying inconsistencies between documentation and code is hard to do, as documentation is written with natural language and code is written with programming languages. 
Rather than considering a limited set of inconsistency patterns in a single programming language~\cite{tan2012tcomment,tan2007icomment,xu2025multiparampython,zhang2024rustc4}, we seek to \emph{explore the range} of developer-written inconsistencies that exist in practice, across multiple programming languages.
Rather than treating inconsistency detection as a binary classification problem~\cite{huang2025your, xu2024code, tan2025just, panthaplackel2021deep, rabbi2020detecting, 
cui2025seocd,
igbomezie2024simplicity, 
liu2018automatic,rong2025carllama},
we seek to (1) \emph{find} conflicting fragments in the code and/or documentation and (2) explain the inconsistencies between them.
Given that it is often difficult to determine whether the code or the documentation is buggy, we believe this exposition could help 
inform developers of potential inconsistencies and promote further fixes.


Pre-trained LLMs enable generic code-documentation inconsistency detection. 
Unfortunately, we found that when asked to identify code-documentation inconsistencies under standard prompting techniques, LLMs have unacceptably high flag rates (i.e., the number of tested functions flagged as inconsistent). 
Not only does the open-source model LLaMA3.1-70B achieve a flag rate of 90-97\%, but the proprietary model GPT4.1 also reaches 82-91\% on our in-the-wild datasets of 1,991 functions from 20 real-world codebases across Python, TypeScript, C++, and Java. 
Such high flag rates are undesirable, as they could lead to alert fatigue and the abandonment of the analysis tool~\cite{sadowski2015tricorder}. Moreover, we observe that a substantial proportion of the identified inconsistencies arise from LLMs eagerly pointing out that some source code details are undocumented. 
We call such inconsistencies \emph{under-promises}. 
Under-promises are \emph{incompleteness} inconsistencies that reflect the natural tendency of documentation to be more concise than code. 
Surfacing all of these would not only obscure the \emph{incorrectness} issues we seek to surface, but also result in a flag rate so high (>90\%) that it becomes untractable for human analysis.

We introduce \toolname{}, a lightweight, multi-language code-documentation inconsistency detector, with the integration of our \emph{Local Categorization and External Filtering} (LCEF) methodology. LCEF guides LLMs to categorize different types of inconsistencies, with the goal of surfacing incorrectness inconsistencies. \emph{Local Categorization} turns the inconsistency detection into a local completion problem, rather than a long-term reasoning problem, and leverages the chain-of-thought approach to reduce spurious outputs.
In addition, \toolname{} includes a front-end that highlights the detected conflicting fragments in the code and/or documentation for each tested code-documentation pair, along with corresponding explanations.

In our evaluation, we run \toolname{} 
on 1,991 developer-written code-documentation pairs \emph{as they exist} in the wild. This allows us to evaluate \toolname{}'s real-world applicability. These functions are collected from 20 real-world open-source codebases across four programming languages: Python, TypeScript, C++, and Java. 
We find that \toolname{}, using the LCEF methodology, is highly effective at surfacing incorrectness inconsistencies: over 96\% of surfaced inconsistencies are incorrectness ones. 
\toolname{} also greatly reduces the number of functions flagged inconsistent with their documentation: decreasing the flag rate from 90-97\% to 14-19\%. This makes our tool's output much more tractable for developer usage.



Though \toolname{}'s goal is to surface incorrectness inconsistencies, 
our ablation study shows it is effective at increasing the overall precision, from 0.14 to 0.71. 
Due to the well-documented noisiness~\cite{xu2024code,rong2025carllama,panthaplackel2021deep} in the labels of existing automatically labelled dataset~\cite{panthaplackel2021deep}, we manually label the results in our datasets.
Through our manual labelling, we categorize  \toolname{}'s correctly and incorrectly identified inconsistencies. 
This manual analysis also allowed us to confirm the presence of incorrectness inconsistencies in real-world software systems, even those that are popular and well-maintained. 
As such, our evaluation provides a conservative lower bound on code-documentation inconsistencies in practice. 

\rev{In addition, we compare \toolname{} with C4RLLaMA, a state-of-the-art tool. C4RLLaMA outperforms prior work on the Panthaplackel dataset~\cite{panthaplackel2021deep}, which} is constructed from synthetically paired (cross-commit) code and documentation. \rev{On this dataset, \toolname{} achieves a precision of 0.78 while C4RLLaMA achieves a precision of 0.83.}
\rev{On our Java dataset, which we refer to as the \toolname{} dataset, \toolname{}'s precision 
drops slightly from 0.78 to 
0.58. 
In contrast, C4RLLaMA's precision drops 
substantially from 0.83 
to 0.08.}
This suggests a distribution shift between their dataset and ours, which contains code-documentation pairs that occur together in the same commit.

In short, our contributions are as follows: 

\begin{itemize}
    \item The Local Categorization-External Filtering (LCEF) approach 
    to reliably surface  \emph{incorrectness} inconsistencies, 
    while maintaining a reasonable (i.e., significantly below 100\%) flag rate.
    \item 
    The \toolname{} tool, a lightweight, multi-language inconsistency detector that implements LCEF and a frontend to surface generic inconsistencies. 
    \toolname{} highlights conflicting code/documentation snippets along with explanations. 

    \item 
    We observe different 
     performance by the state-of-the-art, C4RLLaMA~\cite{rong2025carllama}, on \rev{the \toolname{} dataset} versus the \rev{Panthaplackel dataset}~\cite{panthaplackel2021deep}. 
    This suggests future work evaluating on such synthetically created inconsistencies should also evaluate on
    naturally occurring ones 
    in real-world codebases. 
    
    \item A conservative lower bound on incorrectness issues between functions and their docs. 
    Through our manual validation of \toolname{} results, we found incorrectness inconsistencies in 11\% of examined functions, across 20 projects in 4 programming languages. 
\end{itemize}




\section{Related Work and Design Decisions}
\label{sec:related}
We begin by contextualizing how current literature motivates our
 design and evaluation decisions.

\subsection{Generic vs Specific Inconsistency Detection}
\label{subsec: inconsistency_detection}
In building a code-documentation inconsistency detection tool, a decision must be made about what kinds of inconsistencies should be detected. 
Some approaches choose to focus only on a few types of inconsistency which can be mapped to particular natural language patterns (\emph{pattern-based}). 
Others attempt to use more sophisticated natural language analysis to surface \emph{generic} inconsistencies. 

\textit{Pattern-Based Inconsistency Detection.} As documentation is a natural language artifact, handling it in a precise manner requires careful use of NLP methods. One solution is to focus on analyzing a few natural language patterns whose meaning can be precisely understood.  
icomment~\cite{tan2007icomment} looks for rule-containing documentation by 
checking for
imperative words (e.g., \textit{should}, \textit{must}), and filters this down to lock-related and call-related rules with other keywords 
(e.g., \textit{lock}, \textit{acquire}, \textit{release}). It achieves a precision of 0.61 on a C/C++ dataset. tcomment~\cite{tan2012tcomment} uses the structure of javadocs to infer rules about whether specific parameters are nullable, and checks these inferred rules against dynamic behaviour. It achieves a precision of 0.48 on its Java dataset. 


Even with LLMs, 
tools 
may 
focus on particular patterns of inconsistencies.  \textsc{MPChecker}~\cite{xu2025multiparampython} focuses on \emph{multi-parameter constraint} inconsistencies between API documentation and code. The approach relies on symbolic execution to extract code inconsistencies, making it hard to apply beyond its Python context. 
RustC4 \cite{zhang2024rustc4} uses LLMs to extract an integer range, collection emptiness, and index constraints from documentation. 
It compares these to constraints extracted by a static analysis pass over Rust ASTs. Both of these approaches achieve very high precision ($>0.95$), but remain restricted to particular patterns of inconsistencies in particular programming languages.  

\textit{Generic Inconsistency Detection.} 
A number of works use machine learning to identify arbitrary code-doc inconsistencies. In theory, they could find any semantic difference between code and its documentation.
Most learning-based approaches~\cite{huang2025your, xu2024code, tan2025just, panthaplackel2021deep, rabbi2020detecting, 
cui2025seocd,
igbomezie2024simplicity, 
liu2018automatic,rong2025carllama} 
treat inconsistency detection between code and comments as a binary classification problem. 
They train~\cite{rabbi2020detecting,panthaplackel2021deep,tan2025just,huang2025your,igbomezie2024simplicity, cui2025seocd, liu2018automatic} or fine-tune~\cite{rong2025carllama} machine learning models to achieve better performance across metrics including accuracy and F1 score.

To the best of our knowledge, C4RLLaMA~\cite{rong2025carllama} is the state-of-the-art (SOTA) in both \textit{just-in-time} and \textit{post-hoc} inconsistency detection, outperforming relevant prior works on a publicly available Java dataset~\cite{panthaplackel2021deep}. 
C4RLLaMA is a fine-tuned LLM based on CodeLLaMA that performs binary inconsistency detection between code and documentation.
It also provides rectifications (i.e., fixed versions of the documentation) when the binary label is inconsistent. 



\textbf{Design Choice:} 
Our goal is to detect inconsistency errors developers introduce into real-world software systems; we thus focus on \emph{generic} inconsistency detection.
Our aim is similar to prior generic detection approaches~\cite{huang2025your, xu2024code, tan2025just, panthaplackel2021deep, rabbi2020detecting, 
cui2025seocd,
igbomezie2024simplicity, 
liu2018automatic,rong2025carllama}, in that we also rely on machine learning 
manage the gap between natural language and source code.
This allows us to find generic inconsistencies, but 
trades-off precision for generality 
compared to the pattern-based approaches. 
Unlike most of prior work, we do not rely on a pre-labelled dataset, but instead use manual labelling to evaluate our tool. 

\subsection{Timing of  Inconsistency Detection}

Inconsistency detection between code and documentation can be conducted at two different cadences: \textit{Just-In-Time} and \textit{Post Hoc}~\cite{panthaplackel2021deep}. \textit{Just-In-Time} means the tool is run when the code or documentation are being modified. 
This provides extra input to a detection approach beyond the code and its documentation: the \textit{diff} describing how the code or documentation has changed. 
\textit{Post Hoc} approaches, on the other hand, perform inconsistency detection at some other time, lacking information about a specific \textit{diff} or commit.

\textbf{Design Choice:} To maximize flexibility, we want \toolname{} to apply in as many development situations as possible.
By supporting \textit{Post Hoc} execution, \toolname{} does not need to be tied to a specific \textit{diff}. 
This means \toolname{} can analyze code repositories in the wild, where inconsistencies may have been committed, without having to look through commit history. 
Thus \toolname{} is not 
comparable to the \textit{Just-In-Time} approaches~\cite{stulova2020towards,liu2021just,xu2024code,tan2025just,huang2025your} which require some representation of the \textit{diff} to work. Orthogonally, most \textit{Just-In-Time} approaches are  built and evaluated for Java~\cite{stulova2020towards,liu2021just,xu2024code,tan2025just}, with one approach also evaluated on Python~\cite{huang2025your}. 
We evaluate \toolname{} on Python, TypeScript, Java, and C++.

\subsection{Surfacing Detected Inconsistencies}

What does it mean to detect an inconsistency? Many prior works on generic inconsistency detection treat inconsistency detection as a binary classification problem~\cite{huang2025your, xu2024code, tan2025just, panthaplackel2021deep, rabbi2020detecting, 
cui2025seocd,
igbomezie2024simplicity, 
liu2018automatic}. Thus, the only thing these tools can surface to developers is a binary label: \emph{``this function is inconsistent with its documentation''}. A single binary label may be sufficient for a developer analyzing a shorter method, such as the one in Fig.~\ref{fig:direct-mismatch-example}. However, once methods are longer, it is much harder to pinpoint, from a single binary judgment, what inconsistency has been detected. We aim to provide more information to developers, so they can quickly check whether the detected inconsistency is valid or not. 

Some approaches aim to both classify and
\emph{automatically rectify} inconsistencies in the tested code or comments~\cite{liu2021just, rong2025carllama}.
This gives much more information to the user about the detected inconsistency. If the user compares the original documentation and the rectified documentation, they can try to back-infer the detected inconsistency. But automated documentation rectification assumes that documentation is always the incorrect artifact. We argue that an incorrectness inconsistency likely indicates an error, but we \emph{cannot} know if it is an error in the \emph{documentation} or the \emph{code}. Consider Fig.~\ref{fig:direct-mismatch-example}: without detailed knowledge about the codebase, we cannot know whether the code is missing an addition, or the documentation is an incorrect specification. 

\textbf{Design Choice:} Because we cannot know whether code or documentation should be rectified, and current tooling is not accurate enough to fully automate rectification ~\cite{liu2021just, rong2025carllama}, we believe it is essential to keep developers in the loop. 
Instead of automatically fixing inconsistencies, \toolname{} \emph{exposes and explains} them. 
Given a function and its documentation, \toolname{} outputs an HTML page which highlights inconsistent code and documentation snippets and provides an explanation. Fig.~\ref{fig:over-promise-example},~\ref{fig:direct-mismatch-example} and~\ref{fig:underpromise-example2} are all examples of \toolname{} output.
We believe this output format allows developers to be clearly informed about the nature of the inconsistency so they can fix the code or documentation accordingly.
This output also improves our own ability to accurately manually label any given result.

\subsection{Data Labelling and Contamination}
\label{sec:rw-labelling}
Manually labelling inconsistencies is hard. 
But training machine learning models requires plentiful labelled data. 
Several recent approaches~\cite{xu2024code,rong2025carllama,cui2025seocd} use Panthaplackel et. al's~\cite{panthaplackel2021deep} dataset. The procedure to collect this data  takes two consecutive versions of  a (documentation, code) 
pair: \textlangle$D_1,C_1$\textrangle~and \textlangle$D_2,C_2$\textrangle. Then, it labels the pair \textlangle$D_1,C_2$\textrangle~as \emph{inconsistent} if the code \emph{and} documentation changed between versions (i.e., $D_1\neq D_2, C_1\neq C_2$), and labels the pair \textlangle$D_1,C_2$\textrangle~as \emph{consistent} if only the code changes (i.e., $D_1=D_2, C_1\neq C_2$). This falsely labels pairs as consistent when code changes in a semantically meaningful way, but documentation fails to be updated. 
Further, it has been pointed out that if documentation changes in a semantics-preserving manner, \textlangle$D_1,C_2$\textrangle~will be falsely labelled as inconsistent~\cite{xu2024code}.
Finally, the entire set of inconsistencies are doc-code pairs which never actually occur together in codebases (\textlangle$D_1,C_2$\textrangle~does not occur in the codebase if $D_1\neq D_2$). This introduces potential distribution shift between these synthetic datasets and code in the wild. 

The labels in this dataset are known to be noisy.
Panthaplackel et. al included a ``clean'' subset, where these labels are validated by humans~\cite{panthaplackel2021deep}. Post Hoc performance on the clean dataset was slightly lower than on the full dataset, with precision in the $0.54-0.61$ range rather than the $0.56-0.63$ range. This clean dataset has been reused in some subsequent work~\cite{xu2024code}, but not all~\cite{cui2025seocd,rong2025carllama}. Recent works included training methodologies to account for noise in the dataset labels~\cite{xu2024code,rong2025carllama}. As part of creating this cleaned dataset, Panthaplackel et al.~\cite{panthaplackel2021deep} found that labels were wrong for 26-28\% of \emph{inconsistent} examples, and  8-12\% of \emph{consistent} examples. If the sample analyzed for the clean dataset creation is representative of the rest of the sample, this is a large amount of noise.


There is an additional threat with any procedure for collecting labelled data which relies on historical information. As LLMs are trained on a huge amount of code, if we rely on historically rectified documentation issues to create a dataset, this may lead to data contamination between the LLM's training set and our test set.

\textbf{Design Choice:} Due to these shortcomings, we choose to run our tool on code and documentation as it exists in-the-wild. In particular, we curate functions (or methods) with existing documentation, and run our tool on these pairs. We believe this makes our evaluation less vulnerable to dataset contamination, as this should be the final state of documentation in the LLM's training set. 
However, it means we need to manually analyze the results of our tool to label them as correct or not. 

Manual labelling allows us to categorize reasons for our approach's strengths and weaknesses. But, it means judgments are subject to human error, and inherently based on the opinion of one (for most of our Python, TypeScript, Java, and C++ datasets) or two (for our ablation dataset) computer scientists. 
Nevertheless, we think this evaluation remains representative of how \toolname{} would be used in practice: its outputs would be examined by a developer, who would then decide whether the inconsistency is relevant or not.
\section{Code-Doc Inconsistency Categories}
\label{sec:motivation}

\toolname{} aims to identify code-documentation inconsistencies. That is, our focus is only on inconsistencies that can be spotted from analyzing the source code and documentation of a method. There may be issues in documentation beyond such inconsistencies, such as missing information about dependencies~\cite{aghajani2019software}, or information about temporal relationships between methods~\cite{uddin2015api}. These issues require integrating knowledge about a codebase \emph{beyond} the source code of a method into documentation. Following prior work~\cite{aghajani2019software}, we do not consider these a form of code-documentation inconsistency.

We separate code-documentation inconsistencies into three major categories: over-promises, direct mismatches, and under-promises.
We aim to identify over-promises and direct mismatches, which are \emph{incorrectness} inconsistencies. These are likely to indicate an error in either the documentation or the code. We believe these are higher priority to surface than under-promises, which are \emph{incompleteness} inconsistencies.


\subsection{Over-Promise}
Over-promises are a type of code-documentation inconsistency which point to some part of the documentation not being implemented. 
These correspond to the \emph{``Behaviour described in documentation but not implemented''} subcategory of code-documentation inconsistency described by Aghajani et al.~\cite{aghajani2019software}. Fig.~\ref{fig:over-promise-example} shows a typical example of an \textit{over-promise} found by \toolname{} in the Python pandas project.\footnote{Examples are identified as \texttt{codebase-n}, corresponding to our $n^{th}$ function from the codebase \texttt{codebase}
in our artifact.
} 
The documentation states that the function should ``keep track of whether numexpr was used by storing `True' for each successful use of evaluate with numexpr''. However, this feature is not implemented. 
Over-promises are \emph{incorrectness} issues because functionality developers expect based on the documentation is not present in the implementation. 

\begin{codefigure}
\captionsetup{type=figure, aboveskip=2pt}
\begin{lstlisting}[
  language=Python,
  breaklines=true,
  breakatwhitespace=true,
  xleftmargin=1.5em
]
def set_test_mode(v: bool = True) -> None:
    """Keeps track of whether numexpr was used.(*@\colorbox{lightgray}{Stores an additional `True' for every successful use of}@*)
    (*@\colorbox{lightgray}{evaluate with numexpr since the last `get\_test\_result'.}@*) """
    global _TEST_MODE, _TEST_RESULT
    _TEST_MODE = v
    _TEST_RESULT = []
\end{lstlisting}
\centering
\fbox{\parbox{\dimexpr\columnwidth-2\fboxsep-2\fboxrule\relax}{\footnotesize \textbf{Explanation generated by \toolname{}}: \textit{``The code does not implement storing a True value for every successful use of evaluate with numexpr. It only sets test mode and resets test result.''}}}
\caption{
\small A Python over-promise identified by \toolname{}, along with its highlighting and rationale (\texttt{Pandas-48}).
}
\label{fig:over-promise-example}
\vspace{-10pt}
\end{codefigure}
\vspace{-8pt}

\subsection{Direct Mismatch}
Direct mismatches are code-documentation inconsistencies which point to some part of the documentation logically conflicting with some part of the code. 
These are akin to the \emph{``Code must change to match doc''} subcategory of code-documentation inconsistency described by Aghajani et al.~\cite{aghajani2019software}. However, we do not make a value judgment as to whether it is the code or documentation that must be changed in this case. Fig.~\ref{fig:direct-mismatch-example} shows an example of a \textit{direct mismatch} identified by \toolname{} in the TypeScript Hoppscotch project.\footnote{\url{https://github.com/hoppscotch/hoppscotch}}
The documentation states that the return value should be ``the count returned is highest \texttt{Order}-\texttt{Index} plus 1''. 
However, the code simply returns \texttt{rootCollectionCount[0].orderIndex} without adding one to it. 
These mismatches are \emph{incorrectness} issues, because there is a contradiction between what the implementation does and what its documentation says it will do.

\noindent
\begin{minipage}[t]{0.48\columnwidth}
\begin{figure}[H]
\vspace{-10pt}
\captionsetup{type=figure,aboveskip=2pt}
\centering

\begin{lstlisting}[language=TypeScript,xleftmargin=1.5em]
/**
 * Returns the count of root collections present for a given teamID.(*@\colorbox{lightgray}{The count returned is highest OrderIndex + 1}@*)
 * @param teamID The Team ID
 * @returns Number of Root Collections
 */
private async getRootCollectionsCount(teamID: string) {
    const rootCollectionCount =
      await this.prisma.teamCollection.findMany({
        where: { teamID, parentID: null },
        orderBy: { orderIndex: 'desc' },
      });
    if (!rootCollectionCount.length) return 0;
    return (*@\colorbox{lightgray}{rootCollectionCount[0].orderIndex;}@*);
}
\end{lstlisting}

\fbox{\parbox{\dimexpr\linewidth-2\fboxsep-2\fboxrule\relax}{\footnotesize
\textbf{Explanation generated by \toolname{}}:
\textit{``The code returns the orderIndex as is, without adding 1 to it.''}
}}

\caption{\footnotesize A TypeScript \toolname{}-identified direct mismatch, including highlighting and rationale (\texttt{Hoppscotch-37}).}
\label{fig:direct-mismatch-example}
\end{figure}
\end{minipage}
\hfill
\begin{minipage}[t]{0.48\columnwidth}
\begin{figure}[H]
\vspace{-10pt}
\captionsetup{type=figure,aboveskip=2pt}
\centering

\begin{lstlisting}[language=TypeScript,breaklines=true,breakatwhitespace=true]
/**
 * @name SORTED_FLOATS::operator[]
 * Return the floating point value of the given index into the list.
 */
float SORTED_FLOATS::operator[](int32_t index) {
  (*@\colorbox{lightgray}{it.move\_to\_first();}@*)
  return it.data_relative(index)->entry;
}
\end{lstlisting}

\fbox{\parbox{\dimexpr\columnwidth-2\fboxsep-2\fboxrule\relax}{\footnotesize
\textbf{Explanation generated by \toolname{}}:
\textit{``The documentation does not mention moving to the first element before accessing the data.''}
}}

\caption{\small A C++ under-promise identified by \toolname{}, including highlighting and rationale (\texttt{Tesseract-89}).}
\label{fig:underpromise-example2}
\end{figure}
\end{minipage}


\subsection{Under-Promise}
\label{subsubsec:under-promise}
Finally, under-promises inconsistencies point to some part of the code not being documented. 
These do not correspond to any type of code-documentation inconsistency in Aghajani et al.~\cite{aghajani2019software}; i.e., Aghajani et al. do not characterize these as inconsistencies at all.
Fig.~\ref{fig:underpromise-example2} shows an example of an under-promise in the C++ Tesseract project.\footnote{\url{https://github.com/tesseract-ocr/tesseract}}
This function's documentation does not describe that an iterator is moved to the beginning of the list before returning the value at a specific index in the list. 
We argue that this is an implementation detail a developer using this method would not need to know. So, it is not an incorrectness issue. Depending on a project's documentation practices, some under-promises may be viewed as \emph{incompleteness} issues. But because they are not \emph{incorrectness} issues,  \toolname{} aims to \emph{not} surface under-promises. 


\section{\toolname{}} 
\label{sec:approach}


\begin{figure*}[h!]
\vspace{-10pt}
    \centering
    \includegraphics[width=\linewidth]{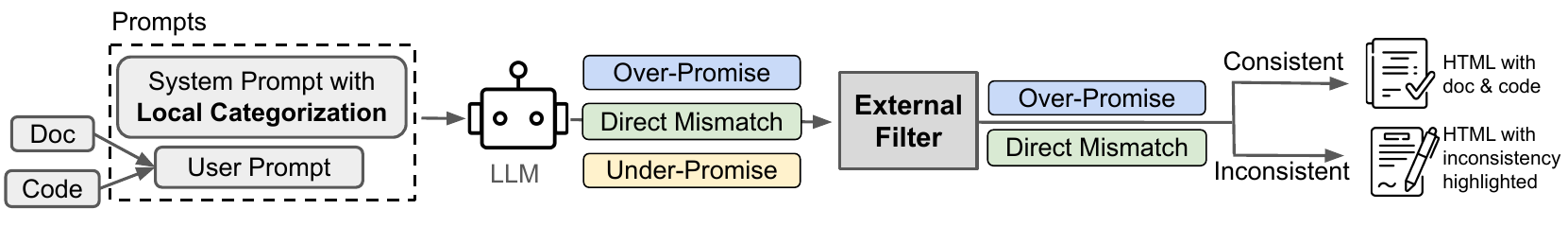}
   \captionsetup{aboveskip=3pt}
    \caption{\toolname{}'s workflow for detecting incorrectness code-doc inconsistencies.
    } 
    \label{fig:workflow-figure}
\vspace{-15pt}
\end{figure*}

We now describe  \toolname{}. As shown in Fig.~\ref{fig:workflow-figure}, given as input 
a function and its 
documentation, \toolname{} first builds system and user prompts, in which it applies our \textit{local categorization}  approach (\S~\ref{subsec:prompts}). Second, it feeds these prompts to the LLM for inconsistency detection (\S~\ref{subsec:inconsistency-detection}). Our prompt design leads the LLM to implicitly categorize inconsistencies into the three categories from Section~\ref{sec:motivation}, and output them in JSON. Third, \toolname{} performs \textit{external filtering} to remove \textit{under-promises}, and generates an HTML page reporting inconsistencies in the other two categories (\S~\ref{subsec:external-filter}).


\subsection{Prompts} \label{subsec:prompts} 
Our prompts are brief by design.
The system prompt consists of two parts: a prefix that introduces the task and input format, and a JSON schema that specifies the output format. The user prompt contains a pair of a function and its method-level documentation.  
As such, the LLM performs the inconsistency detection task in a zero-shot manner.

\subsubsection{System Prompt - Prefix} \label{subsubcsec: prefix} The prefix 
defines the LLM's role, specifies the task, and indicates the input format. The full prefix is as follows:

\begin{lstlisting}[language=text,breaklines=true,breakatwhitespace=true,xleftmargin=2pt,xrightmargin=2pt]
You are a code review expert for the {project_name}{"library"/"framework"/"project"}, working on identifying 
any inconsistencies between function-level code and its documentation. Given the code (enclosed within 
[CODE] [/CODE]) and its accompanying documentation (enclosed within [DOCUMENTATION][/DOCUMENTATION]), please 
identify any inconsistencies or information mismatches between them.
\end{lstlisting}
\vspace{-10px}


\subsubsection{System Prompt - JSON Schema}
\label{subsubsec:system-prompt-json-schema}
We break down the task of code-documentation inconsistency detection 
into five subtasks, similar to a chain-of-thought approach \cite{wei2022chain}. The first two subtasks require the LLM to generate summaries of  (1) the input documentation and (2) the code. These 
tasks aim to prime 
the LLM's understanding of the code and documentation being analyzed. 
The final three subtasks 
instruct the LLM to identify inconsistencies in each category (ref. Section \ref{sec:motivation}): (3) \textit{over-promise}, (4) \textit{direct mismatch}, and (5) \textit{under-promise}. 
Instead of explicitly stating the five reasoning subtasks in the prompt, we encode them 
into a predefined JSON schema with eight keys. 


\textbf{Summary Subtasks.} \emph{(Subtasks 1-2).} The first and second JSON keys, {\small \texttt{"Documentation\_Summary":...}} and  {\small\texttt{"Code\_Summary":...}}, 
prompt the LLM to summarize the documentation and the code. We 
include the word \texttt{Summary} in both keys to emphasize brevity, aiming to reduce 
hallucinated content. 

\textbf{Local Categorization Subtasks.}
The remaining three pairs of JSON keys apply the approach of \textit{local categorization}, which \textit{locally} and \textit{implicitly} guides the LLM to identify {over-promises}, {direct mismatches}, and {under-promises}. Each pair has a check-in key (\textbf{CK}) and a follow-up key (\textbf{FK}). The check-in key is a 
yes/no question tailored to detecting inconsistencies in each category. If the LLM's answer to the check-in-key indicates the presence of inconsistencies in the category, the 
follow-up key prompts the LLM to explain these inconsistencies.

In this manner, the task of identifying inconsistencies between code and documentation, and the category of these inconsistencies, is reduced to a \emph{local completion} problem.
The LLM needs not ``remember'' the definitions of each inconsistency category; 
it needs only complete the output format. As the LLM sequentially generates outputs for each pair of keys, it is implicitly guided to detect and list inconsistencies in each category, if any.

The next pair of JSON keys, \textit{(Subtask 3)}, performs local categorization for \emph{over-promises}. Then the next two keys, \textit{(Subtask 4)}, perform local categorization for \emph{direct mismatches}. For space reasons, we do not include the whole check-in and follow-up keys for these subtasks, but the full prompts can be found in our \href{https://anonymous.4open.science/r/DocCo-E2F0/}{replication package}. 

\textit{(Subtask 5).} The last pair of JSON keys 
prompts the LLM to identify \textit{under-promises}. 
If the answer to the check-in key shown below is ``Yes'', the follow-up key asks it to output undocumented code snippets with explanations.

\begin{center}
\fbox{
\resizebox{0.95\columnwidth}{!}{
\parbox{\columnwidth}{
\begin{itemize}[labelwidth=\widthof{(\textbf{FK}) $\rightarrow$},leftmargin=!]
\footnotesize
\item[(\textbf{CK}) $\rightarrow$]\texttt{"Is\_some\_code\_not\_documented\_or\_mentioned\_in\_the\_documentation?": "Yes" or "No"}
\item[(\textbf{FK}) $\rightarrow$]\texttt{"If\_yes\_to\_Is\_some\_code\_not\_documented\_or\_mentioned\_in\_the\_documentation":
}
\begin{itemize}
    \item\texttt{"original\_code\_snippet\_that\_is\_not\_documented\_or\_mentioned\_in\_the\_documentation"}
    \item\texttt{"explanation"}
  \end{itemize}
\end{itemize}
}
}
}
\end{center}




The follow-up key is verbose; this is by design. 
In the early  experiments, we found that this verbosity was necessary to ensure the LLM's outputs  follow 
the JSON schema.

\subsubsection{User Prompt} Since all detailed instructions are specified in the system prompt, we simply provide a clear test subject through the ``user message''. The following template is used for the user prompt, with placeholders for the documentation and code.

\begin{center}
\fbox{
\parbox{0.65\columnwidth}{
\footnotesize
\noindent\texttt{[DOCUMENTATION]\{doc\}[/DOCUMENTATION]\textbackslash n}
\texttt{
[CODE]\{code\}[/CODE]
}
}
}
\end{center}

\subsection{Inconsistency Detection}
\label{subsec:inconsistency-detection}

For each pair of a function and its 
documentation, \toolname{} first generates the prompts as described 
above. 
To detect inconsistencies, it feeds these prompts to the open-source LLM LLaMA 3.1–70B~\cite{dubey2024llama}. 
The output is in JSON format, following the 
schema specified in the system prompt. In rare cases (3.7\% of evaluated pairs) where the output does not match this format, we cannot extract inconsistencies from the LLM output. Thus \toolname{} does not report any inconsistencies. 
We 
set the temperature to 0 to encourage deterministic results from the LLM.

\subsection{External Filtering}
\label{subsec:external-filter}
\toolname{}'s External Filtering component takes in the JSON object output by the LLM. This component filters out inconsistencies in the third category, \textit{under-promises}, by dropping content under the third check-in key/follow-up key pair (Subtask 5  in \S~\ref{subsubsec:system-prompt-json-schema}). Our HTML backend reports any remaining inconsistencies by highlighting the relevant code/doc, and rendering the LLM-generated explanations. 
The \toolname{} source code and all HTML files, including those for Figures~\ref{fig:over-promise-example}-\ref{fig:underpromise-example2} are included in the artifact (\S~\ref{sec:data_availability}).


\section{Experimental Methodology}
\label{sec:methods}

Our evaluation seeks to provide insight into the kinds of inconsistencies present in real-world developer-written code-documentation pairs and the effectiveness of \toolname{} at uncovering those inconsistencies.
Experimental design decisions all have costs and benefits. Following best-practice advice~\cite{design-tradeoffs}, we report the \textbf{implications} of our design decisions along with our experimental methodology rather than a separate threats to validity section. 

To perform this evaluation, we require access to real-world code and documentation written by developers. 
As discussed in Section~\ref{sec:rw-labelling}, most recent prior work in this area relies on synthetic data that assumes all existing code-documentation pairs in existing code are consistent.
The \textbf{implication} of this choice is that prior datasets do not align with developer beliefs that code-documentation inconsistencies exist in practice~\cite{Lethbridge2003HowSE, tan2007icomment, ratol2017detecting}.
Consequently, we develop an authentic dataset consisting of real-world developer-written code-documentation pairs.

\subsection{Evaluation Setup}
\label{subsec:evaluation_setup}
All evaluations are conducted on an NVIDIA A100 \SI{80}{GB} GPU, the same hardware on which LLaMA 3.1–70B~\cite{dubey2024llama} performs inference. On average, \toolname{} took approximately 22 seconds to evaluate each function.
The \textbf{implication} of this choice is that our approach is primarily evaluated using a single model--- although Section~\ref{sec:rq4} examines whether proprietary model gives different results.

\subsection{Evaluation Datasets}
\label{subsec:benchmark-datasets}

To perform our evaluation, we constructed two datasets. Our primary polylingual dataset is composed of code-documentation pairs from four programming languages and is used to gain insight into \toolname{} and the nature of developer-written code-documentation inconsistencies.
A second single-language ablation dataset is used to gain insight into specific \toolname{} design decisions.


\subsubsection{Primary polylingual dataset (Python, TypeScript,  C++, and Java)} 
\label{subsubsec: wild-datasets} 

To understand how \toolname{} performs in detecting incorrectness inconsistencies across multiple programming languages,
we built a dataset including 20 projects, aiming for 100 documented functions per project. 5 projects were written in each of four programming languages: Python, TypeScript, C++, and Java.
We selected 5 popular open-source projects\footnote{Details of the 20 projects are included in the linked artifact.} for each language by the following protocol: 

\begin{enumerate}
\item Sorting by the most starred projects on GitHub. 
\item The project must have at least two years of development history. 
\item The project must be under active development, with at least 5 issues or pull requests closed in the last 2 months.  
\item At least 60\% of the project's source code is written in the target programming language, according to the GitHub repository statistics.
\end{enumerate}

%
%

From each of the selected 20 projects, we randomly sampled 100 functions that satisfy the following requirements:
\begin{itemize}
    \item The function has a method-level documentation that follows documentation syntax for the given language (e.g., Javadoc).
    \item The function is implemented in the language of interest.
    \item Both the function and its documentation have more than 7 tokens. 
    \item There are no duplicate code-documentation pairs. 
\end{itemize}

Unfortunately, two projects had fewer than 100 methods meeting these criteria.
Only 94 functions from the TypeScript-based Nest project and 97 functions from the C++-based llama.cpp project met these criteria. 
The final dataset contains 1,991 code-documentation pairs: 500 in Python, 494 in TypeScript, 497 in C++, and 500 in Java. 
The \textbf{implication} of this dataset is that it provides an authentic view into code-documentation pairs as they exist in deployed software systems. This allows us to examine these pairs for real-world inconsistencies.
However, we cannot automatically assign consistent/inconsistent labels to the code-documentation pairs. Since they are in popular production software, we conjecture that \emph{many} of them are consistent, but we cannot know a priori which ones are. 


To address this challenge, we invested substantial effort to manually label a subset of this data.
Determining whether a function and its documentation are consistent is time-consuming work.
To make this process tractable, we labelled only those code-documentation pairs that \toolname{} flagged as inconsistent.
This was a more direct process as the tool provides rationale and explicitly-highlighted evidence from the documentation and code that a labeller can assess as either correct or incorrect.
Labelling a \toolname{}-flagged code-documentation pair takes between 5-15 minutes: longer times are usually because we need to analyze code in the repository outside the core function (i.e., its callees) to determine whether the inconsistency is correct. 
Each code-documentation pair was labelled by a single author.
The \textbf{implication} of this labelling methodology was that we are able to assess \toolname{}'s precision (as all flagged pairs are labelled for true/false positive analysis), but cannot assess recall (as false negatives were not examined). Since only a single labeller examined each output, consistency could be a concern.

To mitigate the single-coder labelling threat, we performed an inter-coder reliability analysis. This evaluates how consistently different labellers would assess the same code-documentation pairs.
Two authors independently labelled \rev{200} randomly chosen flagged functions, \rev{50 from each of four programming languages: Python, TypeScript, Java, and C++}.
The Cohen's Kappa for these labels is \rev{0.737}, which is classified as \textit{\rev{substantial}} agreement~\cite{kappa}. The \textbf{implication} of this is that individual labellers can reliably determine whether \toolname{}-flagged code-documentation pairs in our dataset are inconsistent.

\subsubsection{Ablation dataset (Python)}
\label{subsubsec:python-dataset}
Our ablation dataset contains 52 fully labelled code-documentation pairs in Python.
These pairs were randomly selected from two popular Python projects.\footnote{Pandas: \url{https://github.com/pandas-dev/pandas}; Requests: \url{https://github.com/psf/requests}}
The 52 functions were chosen using the same sampling methodology described in the primary dataset.
Unlike the primary dataset, we labelled all 52 functions \emph{a priori}: \toolname{} did not classify them first.
Additionally, all 52 functions were independently labelled by two authors.
The code-documentation pairs are labelled as either consistent or inconsistent; when inconsistent, the source of the inconsistency is noted. 
Any differences in labels were discussed among both coders until agreement was reached for all 52 code-documentation pairs.
The \textbf{implication} of this dataset is that this fully-labelled dataset allows us to investigate \toolname{} in greater detail by measuring precision, recall, accuracy, and F1 score across ablation variants. 

\subsection{Evaluation Metrics} 
\label{subsec:evaluation-metrics}
Our metrics have two levels of granularity. 
\textit{Function-level} metrics evaluate \toolname{}'s 
on a per-function basis. \textit{Inconsistency-level} metrics evaluate each identified inconsistency separately.



\subsubsection{Function-Level Terminology}
\label{function-level-term}


\begin{enumerate}[leftmargin=2em]
    \item \textbf{True Positive (TP):} 
    The function is inconsistent with its documentation, and \toolname{} surfaces at least one of the 
 inconsistencies. 
    \item \textbf{False Positive (FP):} The function is consistent with its  documentation, but \toolname{} surfaces at least one inconsistency. 
    \item \textbf{True Negative (TN):} The function is consistent with its  documentation, and \toolname{}  surfaces no inconsistencies. 
    \item \textbf{False Negative (FN):} The function is inconsistent with its  documentation, but \toolname{}  surfaces no inconsistencies.
    \item \textbf{Flag Rate:} The percentage of functions flagged as inconsistent with their method-level docs.
    \begin{equation*}
\small
\text{Flag Rate} = \frac{\text{\# Functions Flagged as Inconsistent}}{\text{\# Total Tested Functions}}
\end{equation*} 
\end{enumerate} 

\subsubsection{Inconsistency-Level Terminology}
\begin{enumerate}[leftmargin=2em]
    \item \textbf{True Positive (TP):} The reported inconsistency is indeed an inconsistency between code and documentation.
    \item \textbf{False Positive (FP):} The reported inconsistency is wrong. 
    \item \textbf{Under-promise Rate (U.P. Rate):} The percentage of \textit{under-promises}  in the final output.
    \begin{equation*}
\small
\text{Under-Promise Rate} = \frac{\text{\# Under-Promises}}{\text{\# Total Reported Inconsistencies}}
\end{equation*}
\end{enumerate} 

\subsubsection{\rev{C4RLLaMA Function-Level Terminology}}
\label{sec:function-level-carllama}
\rev{
When
C4RLLaMA~\cite{rong2025carllama} flags a code-doc pair as inconsistent, it does not explain the inconsistency, but directly outputs rectified documentation. Labelling C4RLLaMA on whether its rectification is correct would be more strict than our true positive above. To align our measurement as much as possible with our function-level metrics, we use the following definitions of function-level true- and false-positive: 
} 
\rev{\begin{enumerate}[leftmargin=2em]
    \item \textbf{True Positive (TP):} 
    C4RLLaMA identifies the function as inconsistent with its documentation, and updates at least one documentation fragment where an 
    inconsistency occurs.  Note it need \emph{not} correctly rectify the inconsistency.
    \item \textbf{False Positive (FP):} 
    C4RLLaMA identifies the function as inconsistent with its documentation, but does not update any documentation fragment in which there is an 
    inconsistency.
\end{enumerate} 
}
\noindent\rev{For both \toolname{} and C4RLLaMA, we consider only \emph{incorrectness} inconsistencies in the definitions.}







\section{Results}
\label{sec:eval}

We aim to answer the following questions: 
\begin{enumerate}[leftmargin=*, label=\textbf{RQ\arabic*.}]
    \item How does \toolname{} compare with the state-of-the-art C4RLLaMA?
    \item How does \toolname{} perform on real-world code-doc inconsistencies across multiple programming languages? What are its sources of false positives?
    \item How prevalent are code-doc incorrectness inconsistencies in practice? What kinds of incorrectness inconsistency does \toolname{} find?
    \item  How do our  design decisions, in particular \textit{Local Categorization and Exernal Filtering} (LCEF), affect \toolname{}'s performance?
    \item  How does LCEF generalize to other LLMs? 
\end{enumerate}

\subsection{RQ1: \toolname{} vs. State-of-the-Art}
\label{sec:rq1}

As mentioned in Section~\ref{subsec: inconsistency_detection}, C4RLLaMA~\cite{rong2025carllama} is the state-of-the-art (SOTA) in both \textit{just-in-time} and \textit{post-hoc} inconsistency detection on \rev{the Panthaplackel dataset~\cite{panthaplackel2021deep}, in Java}.  All of the dataset's inconsistent examples are code-comment pairs that never appear together in the real codebase \rev{(ref. Section~\ref{sec:rw-labelling}), i.e., they are synthetic.}
\rev{The setting we target in this work is} the \textit{post-hoc} setting. 
We used the scripts provided in the C4RLLaMA replication package to fine-tune CodeLLaMA. \rev{We use the evaluation setup described in Section~\ref{subsec:evaluation_setup}.} Detailed outputs can be found in our artifact.

\subsubsection{\rev{\toolname{} Dataset: Flag Rates}}
We compare \toolname{} with C4RLLaMA on our Java dataset, which contains 500 code-documentation pairs as they appear in real-world codebases (ref. Section~\ref{subsec:benchmark-datasets}) \rev{We refer to it as the \toolname{} dataset.} We summarize \toolname{} and C4RLLaMA's 
flag rates  in Table~\ref{tab:baseline-comparison-flag-rate}. 
The flag rate represents the proportion of functions flagged by a tool as inconsistent with its documentation. 
\toolname{} takes code and the entire documentation block as input, which may contain multiple components (e.g., Summary, @param, @return). \toolname{} flags 92 of 500 real-world code-documentation pairs as inconsistent, an 18\% flag rate. 

The dataset on which 
C4RLLaMA was trained and evaluated, however, separates this full documentation block into individual documentation components (i.e., Summary, @param, @return).
To make our dataset more like this, we extracted the Summary, @param, and @return components (if they exist) from the full documentation, and pair the corresponding source code with each of them. 
This led to a total of 1,159 pairs: 464 \textlangle code, @param only\textrangle, 234 \textlangle code, @return only\textrangle, and 461 \textlangle code, summary only\textrangle. Then, we ran C4RLLaMA on these three types of pairs.





With the documentation split in this way, we see C4RLLaMA flag something as consistent. 
C4RLLaMA flags 463 out of 464 \textlangle code, @param only\textrangle, all 234 \textlangle code, @return only\textrangle\ and all 461 \textlangle code, summary only\textrangle\ pairs as inconsistent. 
But this still leads to a 100\% flag rate on code-documentation pairs as a whole---one of the documentation parts was flagged for each code-documentation pair. 


\begin{table}[h]
\centering
\small
{\renewcommand{\arraystretch}{1}
\captionsetup{belowskip=1pt, aboveskip=5pt} 
\caption{\small Flag rate comparison between \toolname{} and state-of-the-art across four types of pairs on the Java dataset (Section~\ref{subsec:benchmark-datasets}). \textlangle code, \textit{X} only\textrangle\      corresponds to the flag rate for input pairs of code and individual documentation component \textit{X}. \textit{Full documentation} includes all available documentation components. \textit{\%Flagged code-doc pairs} indicates the percentage of code-documentation pairs flagged as inconsistent. }
\label{tab:baseline-comparison-flag-rate}
\scalebox{0.8}{
\setlength{\tabcolsep}{40pt}
\begin{tabular}{@{} l c r}
  \toprule 
   & \textbf{\toolname{}} & \textbf{C4RLLaMA}\hphantom{XX} \\
  \midrule
  \textlangle code, full documentation\textrangle & 18.4\% (92/500) & 100\hphantom{X}\%~(500/500) \\
   \textlangle code, @param only\textrangle & -- & 99.8\% (463/464) \\
   \textlangle code, @return only\textrangle & -- & 100\hphantom{X}\%~(234/234) \\ 
   \textlangle code, summary only\textrangle & -- & 100\hphantom{X}\%~(461/461) \\
   
   \midrule
   \% Flagged code-doc pairs & \cellcolor{gray!20}{18.4\% (92/500)} & 100\hphantom{X}\% (500/500) \\
   \# Flagged inconsistencies & \cellcolor{gray!20}{175} & 500 (full doc) or 1,158 (components) \\
  \bottomrule
\end{tabular}}}
\vspace{-10pt}
\end{table}

\subsubsection{\rev{\toolname{} Dataset: Manual Precision}}
\rev{We further examine C4RLLaMA's precision on this dataset}. \rev{The} 100\% flag rate means we cannot apply our manual labelling methodology \rev{on \emph{all}} of the 1,158 inconsistencies flagged by C4RLLaMA. \rev{Instead, we randomly sampled 50 
code-documentation pairs from the \toolname{} dataset that either \toolname{} or C4RLLaMA flagged as inconsistent. This gives us a total of 50 \textlangle{}code, documentation\textrangle{}, 47 \textlangle{}code, Summary\textrangle{} , 32 \textlangle{}code, @param\textrangle{}, and 24 \textlangle{}code, @return\textrangle{} pairs. We chose a sample size of 50 because labelling C4RLLaMA outputs is more time-consuming than labelling \toolname{} outputs. To reduce sample size threats, we report precision results with a 95\% Wilson score confidence interval~\cite{Dunnigan2008ConfidenceIC}.
} 

\rev{
We label \toolname{} and C4RLLaMA outputs using the true- and false-positive definitions introduced in Section~\ref{function-level-term} and Section~\ref{sec:function-level-carllama} respectively. Two authors independently label the C4RLLaMA outputs, and the remaining authors discuss and resolve any 
labelling disagreements.}

\rev{As shown in Table~\ref{tab:cross-table}, under our measure of precision, C4RLLaMA achieves a precision of 0.08 across all pairs, with a 95\% Wilson score confidence interval of 0.05 -- 0.14. In contrast, \toolname{} achieves a precision of 0.58 [0.47 -- 0.67], based on our manual labels for all flagged pairs (ref. RQ2 in Section~\ref{sec:rq2}). 
A precision of 0.05 -- 0.14 is much lower than the 0.94 reported by C4RLLaMA on the Pathaplackel dataset~\cite{rong2025carllama}. Next, we investigate performance on the Pathaplackel data to determine whether the change in precision is due to the dataset shift or the different measure of precision.} 
\vspace{-8pt}
\begin{table*}[h]
\centering
\rev{
\small
{\renewcommand{\arraystretch}{1}%
\captionsetup{belowskip=1pt, aboveskip=5pt}
\caption{\small \rev{Precision of \toolname{} and C4RLLaMA on the Panthaplackel and \toolname{} datasets under our precision measures. 
 Ranges indicate 95\% confidence intervals~\cite{Dunnigan2008ConfidenceIC}. 
}}
\label{tab:cross-table}
\scalebox{0.8}{
\begin{tabular}{@{\extracolsep{5pt}} l c c @{}}
  \toprule 
    & \textbf{Panthaplackel Dataset} & \textbf{\toolname{} Dataset} 
   \\
  \midrule
   \toolname{}  & 0.78 [0.62 - 0.88] & 0.58 [0.47 - 0.67] \\
    C4RLLaMA  & 0.83 [0.67 - 0.92] & 0.08 [0.05 - 0.14]\\
  \bottomrule
\end{tabular}
}}}
\vspace{-1.5em}
\end{table*}

\subsubsection{\rev{Panthaplackel Dataset and Metrics}}
\rev{We run C4RLLaMA and \toolname{} on the \emph{Panthaplackel~\cite{panthaplackel2021deep} dataset}.} 
\rev{This dataset provides consistent/inconsistent labels: Table~\ref{tab:rq1-synthetic-dataset-performance} presents the results in terms of these labels. We successfully reproduce the C4RLLaMA results reported in their paper~\cite{rong2025carllama}, averaging a 2.7\%\footnote{Results vary across runs, with differences ranging from 1.3\% to 2.7\%.} difference across the metrics, using their scripts and the fine-tuned model.} 
\rev{Overall, C4RLLaMA achieves higher metric values than \toolname{} (e.g., 0.87 vs. 0.67 in precision). However, this does not necessarily indicate that it outperforms \toolname{} on the Panthaplackel dataset. The reasons are twofold. First, human analysis on a subset suggests that 17-20\% of the  labels in this dataset are incorrect~\cite{panthaplackel2021deep}. Second, as the labels are binary, the metric does not measure whether the surfaced inconsistencies are \emph{correct}, just that \emph{an} inconsistency is surfaced.}  

\subsubsection{\rev{Panthaplackel Dataset: Manual Precision}}
\rev{ To answer whether C4RLLaMA's 0.08 precision on our dataset is due to the dataset change or our precision measure, we apply our precision measure to the Panthaplackel Dataset. The Panthaplackel dataset does not contain full \textlangle{}code, documentation\textrangle{} pairs, as \toolname{} expects. We  consider \textlangle{}code, Summary only\textrangle{} pairs, as these are closest to full documentation. Thus, we perform our labelling on 50 randomly-sampled \textlangle{}code, Summary only\textrangle{} pairs that either \toolname{} or C4RLLaMA flagged as inconsistent. Again, two authors independently label the C4RLLaMA and \toolname{} outputs using the true- and false- positive definitions from Sections~\ref{function-level-term} and~\ref{sec:function-level-carllama}. The remaining authors discuss and resolve the 
labelling disagreements. }

\rev{
The left column of Table~\ref{tab:cross-table} shows the results.
Under our measure of precision, C4RLLaMA achieves a precision of 0.83 on the Panthaplackel dataset, (95\% confidence interval~\cite{Dunnigan2008ConfidenceIC} : 0.67 -- 0.92). This is comparable to C4RLLaMA’s precision on the same dataset with its own measure of precision: 0.89.  From this, we conclude that, C4RLLaMA’s precision drop to 0.08 on the \toolname{} dataset is due to the \emph{dataset change} rather than \emph{our precision measure}. In Table~\ref{tab:cross-table}, \toolname{} shows more stable performance than C4RLLaMA across the two datasets. While \toolname{} performs comparably to C4RLLaMA on the Panthaplackel dataset (\toolname{}: 0.78, C4RLLaMA: 0.83), it achieves much higher precision on the \toolname{} dataset (\toolname{}: 0.58, C4RLLaMA: 0.08).}

\rev{
At a higher level, good performance on the Panthaplackel dataset, under both measures of precision, does not translate to higher performance on the \toolname{} dataset. We recommend future work evaluating on synthetically created code-doc pairs also report results code-doc pairs as they exist in real-world codebases. 
}



\begin{table}[t]
\centering
\rev{
\small
{\renewcommand{\arraystretch}{1}%
\captionsetup{belowskip=1pt, aboveskip=5pt}
\caption{\small \rev{Performance of \toolname{} and C4RLLaMA on the Panthaplackel dataset using its binary labels.}}
\label{tab:rq1-synthetic-dataset-performance}
\scalebox{0.8}{
\begin{tabular}{@{\extracolsep{1pt}} l c c c c c c c c c c @{}}
  \toprule
  \multicolumn{1}{c}{} & 
  \multicolumn{5}{c}{\textbf{\toolname{}}} & 
  \multicolumn{5}{c}{\textbf{C4RLLaMA}} \\
  \cmidrule(lr){2-6} \cmidrule(lr){7-11}
   & \textbf{Flag rate} & \textbf{Precision} & \textbf{Recall} & \textbf{ Accuracy} & \textbf{F1} 
   & \textbf{Flag rate} & \textbf{Precision} & \textbf{Recall} & \textbf{Accuracy} & \textbf{F1} \\
  \midrule
   \textlangle{}code, @param only\textrangle{}  & 52\% & 0.62 & 0.64 & 0.63 & 0.63 & 47\% & 0.92 & 0.87 & 0.89 & 0.89 \\
   \textlangle{}code, @return only\textrangle{} & 43\% & 0.7 & 0.6 & 0.67 & 0.65 & 48\% & 0.9 & 0.87 & 0.89 & 0.89 \\
   \textlangle{}code, summary only\textrangle{} & 38\% & 0.68 & 0.51 & 0.63 & 0.58 & 51\% & 0.85 & 0.86 & 0.85 & 0.85 \\
   \midrule
    Total & 44\% & 0.67 & 0.59 & 0.65 & 0.63 & 49\% & 0.89 & 0.87 & 0.88 & 0.88 \\
  \bottomrule
\end{tabular}}}
}
\vspace{-5pt}
\end{table}

\begin{tcolorbox}[takeaway]
\textbf{RQ1 Takeaway.}
\rev{Using our manual precision metrics, \toolname{} performs comparably to C4RLLaMA (precision: 0.78 vs 0.83, flag rate: 44\% vs 49\%) on the Panthaplackel dataset. On our \toolname{} dataset, the tools perform very differently (precision: 0.47-0.67 vs 0.05-0.14, flag rate: 18\% vs 100\%). C4RLLaMA's shift in performance suggesst there is a distribution shift between synthetic cross-commit inconsistencies and naturally-occuring ones.}
\end{tcolorbox}

\subsection{RQ2: Inconsistency Detection Effectiveness}  
\label{sec:rq2}

In RQ1 we examined only inconsistency flag rate. The natural follow-up question is: are the flagged issues true inconsistencies? 
To answer this, we run \toolname{} on  1,991 developer-written code-documentation pairs over four programming languages: Python, TypeScript, C++, and Java (\S~\ref{subsec:benchmark-datasets}). We manually analyze \toolname{}'s detected inconsistencies to determine whether they are correct. Further, we categorize \toolname{}'s  most common sources of false positives.

\subsubsection{Results} 

We manually inspect each inconsistency  \toolname{} reports to determine whether it is a true positive (a correctly identified inconsistency) or a false positive (an invalid inconsistency).\footnote{This manual analysis is detailed in Section~\ref{sec:methods}.} 
From this, we can calculate precision (\S~\ref{subsec:evaluation-metrics}).

\begin{table}[t]
\centering
\small
{\renewcommand{\arraystretch}{1}%
\captionsetup{belowskip=1pt, aboveskip=5pt}
\caption{\small \toolname{} performance across four programming languages: Python, TypeScript, C++, and Java; see Section~\ref{subsec:evaluation-metrics} for explanations of all metrics. 
\textit{Flag rate} is the percentage of functions flagged as inconsistent. \textit{\#Inconsistencies} is the total number of inconsistencies reported by \toolname{}. \textit{U.P. rate} is  \emph{under-promise rate}. 
}
\label{tab:rq2}
\scalebox{0.8}{
\begin{tabular}{@{\extracolsep{1pt}} l c c c c c c c c c c @{}}
  \toprule
  \multicolumn{1}{c}{} & 
  \multicolumn{5}{c}{\textbf{Function-level data}} & 
  \multicolumn{5}{c}{\textbf{Inconsistency-level data}} \\
  \cmidrule(lr){2-6} \cmidrule(lr){7-11}
   & \textbf{\# Functions} & \textbf{Flag rate} & \textbf{\# TP} & \textbf{\# FP} & \textbf{Precision} 
   & \textbf{\# Inconsistencies} & \textbf{U.P. rate} & \textbf{\# TP} & \textbf{\# FP} & \textbf{Precision} \\
  \midrule
   Python & 500 & 19\% & 61 & 32 & 0.66 & 176 & 2\% & 105 & 71 & 0.60 \\
   TypeScript & 494 & 16\% & 57 & 24 & 0.70 & 134 & 5\% & 89 & 45 & 0.66 \\
   C++        & 497 & 14\% & 40 & 31 & 0.56 & 123 & 8\% & 67 & 56 & 0.55 \\
   Java       & 500 & 18\% & 53 & 39 & 0.58 & 175 &3\% & 91 & 84 & 0.52 \\
   \midrule
    Total\textsuperscript{\textdagger}/~Mean & 1,991\textsuperscript{\textdagger}\hphantom{l} & 17\% & 211\textsuperscript{\textdagger}\hphantom{.} & \hphantom{i}126\textsuperscript{\textdagger} & 0.63 & \hphantom{l}608\textsuperscript{\textdagger} & 4\% & 352\textsuperscript{\textdagger} & 256\textsuperscript{\textdagger} & 0.58 \\
  \bottomrule
\end{tabular}}}
\vspace{-5pt}
\end{table}

Table~\ref{tab:rq2} shows the results. We split results into  function-level and inconsistency-level. This is because \toolname{} can return multiple inconsistencies for a single function (see \S~\ref{subsec:evaluation-metrics} for details).

Overall, \toolname{} performs well on  \emph{flag rate} and \emph{under-promise rate} across all four languages. The \textit{under-promise} rate ranges from 2\% to 8\%, showing that \toolname{} mostly surfaces the targeted \emph{incorrectness} inconsistencies. 
The flag rate ranges between 14\% to 19\%. We conjecture this is more 
tractable for human analysis than the 100\% flag rate of state-of-the-art (\S~\ref{sec:rq1}). 
\toolname{} achieves higher precision on TypeScript and Python (0.7, 0.66), than on C++ and Java (0.56, 0.58). 


\subsubsection{Sources of \toolname{} False Positives}

Table~\ref{tab:fp-categorization} summarizes the four main reasons for \toolname{}'s false positives across the examined languages. 

\textbf{Lack of API knowledge (35\%):} 
The LLM is unsure about, or assumes missing, functionality in a callee API, even though the callee API actually implements the desired behaviour.
The \texttt{info()} example from  Fig.~\ref{fig:lack of api knowledge} falls in this category. The documentation states that the bug information should be pretty-printed as JSON. However, the function calls \texttt{info()} to get the bug information. As shown in the explanation box, the LLM cannot determine whether \texttt{info()}  generates bug information, and thus flags this as inconsistent.

\begin{codefigure}
\vspace{-9pt}
\captionsetup{type=figure, skip=3pt}
\begin{lstlisting}[language=Python,breaklines=true,breakatwhitespace=true]
def main():
    """Pretty-print(*@\colorbox{lightgray}{the bug information}@*)as JSON."""
    print(json.dumps((*@\colorbox{lightgray}{info()}@*), sort_keys=True, indent=2))
\end{lstlisting}
\fbox{\parbox{\dimexpr\columnwidth-2\fboxsep-2\fboxrule\relax}{\footnotesize \textbf{Explanation generated by \toolname{}}: \textit{The code does not explicitly handle `bug information'. The info() function is called, but its implementation and relation to bug information are unclear.}}}
\caption{\small An example of a false positive reported by \toolname{} due to \textit{lack of API knowledge} (\texttt{Requests-26}). }
\label{fig:lack of api knowledge}
\vspace{-9pt}
\end{codefigure}

\textbf{Semantics Gap (7\%):} The LLM correctly characterizes the semantics of the documentation and code snippet, but fails to identify that the implementation in the code is an instance of the semantics in the documentation, or vice-versa. 
For example, the documentation of \texttt{n8n-43} mentions a feature of ``expiring the current client'' and the code implements it with \texttt{this.client = undefined;}. The LLM flags this function as inconsistent, saying that \textit{``the documentation mentions expiring the current client, but the code only sets it to undefined''}. The LLM fails to identify that setting to \texttt{undefined} is sufficient to meet the definition of ``expire''.

\textbf{Under-Promise (10\%):} 
The LLM points out that some details are not documented (\S ~\ref{subsubsec:under-promise}).  

\textbf{Incorrect Reasoning (48\%):} The LLM incorrectly characterizes code or documentation semantics, or provides erroneous explanations. This is distinct from  \textit{Semantics Gap}, where the LLM correctly characterizes the semantics of both the code and the documentation, but fails to align them. 
We divide this into four subcategories: \textit{temporal constraints}, \textit{contextual info}, \textit{no domain knowledge}, and \textit{others}. \textit{Temporal constraints} refer to cases where the documentation sets additional requirements regarding the usage of this function, but the LLM misinterprets them as features that need to be implemented in the code. 
The same applies to \textit{contextual info}. 
For example, the documentation of \texttt{Tesseract-48} mentions some surrounding context for the function: \textit{``this routine is designed to be used in concert with the KDWalk routine.''} As this synchronization is not explicitly implemented in the function body, the LLM 
flags this as an inconsistency. 
\textit{No domain knowledge} means that detected inconsistency is wrong 
due to a lack of relevant domain-specific knowledge. 

\begin{table}[t]
\centering
\small
\setlength{\tabcolsep}{2pt}

\begin{minipage}[t]{0.48\linewidth}
\centering
{\renewcommand{\arraystretch}{0.91}
\captionsetup{belowskip=1pt, aboveskip=5pt}
\captionof{table}{\small Manual categorization of \textit{incorrectly} identified inconsistencies surfaced by \toolname{} 
}
\label{tab:fp-categorization}
\resizebox{\linewidth}{!}{
\begin{tabular}{@{\extracolsep{3pt}} l r r r r}
  \toprule 
   & \textbf{Python} & \textbf{TS} & \textbf{C++} & \textbf{Java}\\
  \midrule
   Lack of API Knowledge & \cellcolor{gray!20}{39 (55\%)} & 15 (33\%) & 17 (30\%) & 18 (21\%)\\
   Semantics Gap & 1 (1\%) & 4 (9\%) & 2 (4\%) & 10 (12\%)\\ 
   Under-promise & 4 (6\%) & 7 (16\%) & 10 (18\%) & 5 (6\%)\\
   Incorrect Reasoning & 27 (38\%) & \cellcolor{gray!20}{19 (42\%)} & \cellcolor{gray!20}{27 (48\%)} & \cellcolor{gray!20}{51 (61\%)}\\
   \quad Domain Knowledge & 2 & 3 & 0 & 0 \\
   \quad Temporal Constraints & 4 & 0 & 1 & 6\\
   \quad Contextual Info & 5 & 5 & 1 & 18\\
   \quad Others & 16 & 11 & 25 & 27\\
   \midrule
   Total & 71 & 45 & 56 & 84\\
\bottomrule
\end{tabular}}}
\end{minipage}
\hfill
\begin{minipage}[t]{0.48\linewidth}
\centering
{\renewcommand{\arraystretch}{1}
\captionsetup{belowskip=1pt, aboveskip=5pt}
\captionof{table}{\small Manual categorization of inconsistency errors \textit{correctly} surfaced by \toolname{}.}
\label{tab:tp-categorization}
\resizebox{\linewidth}{!}{
\begin{tabular}{@{\extracolsep{3pt}} l r r r r}
  \toprule 
   & \textbf{Python} & \textbf{TS} & \textbf{C++} & \textbf{Java}\\
  \midrule
   Functionality Mismatch &\cellcolor{gray!20}{ 58 (55\%)} & \cellcolor{gray!20}{42 (47\%)} & \cellcolor{gray!20}{38 (57\%)} & 29 (32\%)\\
   Parameter Mismatch & 7 (7\%) & 5 (6\%) & 2 (3\%) & 11 (12\%)\\ 
   Parameter Type Mismatch & 1 (1\%) & 7 (8\%) & 0 (0\%) & 4 (4\%)\\
   Identifier Mismatch & 0 (0\%) & 9 (10\%) & 3 (5\%) & 2 (2\%)\\
   Return Type Mismatch & 3 (3\%) & 11 (12\%) & 4 (6\%) & 1 (1\%)\\
   Unimplemented Feature & 35 (33\%) & 15 (17\%) & 10 (15\%) & \cellcolor{gray!20}{44 (48\%)} \\
   Unused Parameter/Variable & 1 (1\%) & 0 (0\%) & 10 (15\%) & 0 (0\%)\\
   \midrule
   Total & 105 & 89 & 67 & 91\\
\bottomrule
\end{tabular}}}

\end{minipage}
\end{table}

\begin{tcolorbox}[takeaway]
\textbf{RQ2 Takeaway.} 
\toolname{}
almost always surfaces incorrectness inconsistencies (92\%-97\%). 
It achieves higher precision on TypeScript and Python (0.7, 0.66), than on C++ and Java (0.56, 0.58). 
Most false positives are due to \emph{incorrect reasoning}, then \emph{lack of API knowledge}.
\end{tcolorbox}


\subsection{RQ3: Prevalence of Incorrectness Inconsistencies} 
\label{sec:rq3}

In the process of manually labelling all \toolname{}-flagged inconsistencies for RQ2, we found that
incorrectness inconsistencies do exist in maintained software systems.
Of the 1,991 code-doc pairs on which we ran \toolname{}, we confirmed that there were incorrectness inconsistencies for 211 of them (Table~\ref{tab:rq2}, \textit{\#~TP}). 
This conservatively suggests that there exist errors in the code or documentation of at least 11\% of non-trivial documented methods. 

Table~\ref{tab:tp-categorization} summarizes the seven main sources of code-doc inconsistencies we identified. 
They are:


\textbf{Functionality Mismatch (47\%):} 
Features described in the documentation logically conflict with their implementation. 
Fig.~\ref{fig:direct-mismatch-example} gives an example: while the documentation mentions returning the highest \texttt{OrderIndex + 1}, the code implements returning the highest \texttt{OrderIndex}.

\textbf{Parameter Mismatch (7\%):}
Either some documented parameters are not present in the function signature, or only a subset of the parameters defined in the function signature are documented.

\textbf{Parameter Type Mismatch (3\%):} 
Param. types in the docs do not match those in the signature. 

\textbf{Identifier Mismatch (4\%):} Different identifiers refer to the same object in the code vs. the docs. 

\textbf{Return Type Mismatch (5\%):} 
The function’s 
return type differs from what is in the docs.

\textbf{Unimplemented Feature (30\%):} A feature specified in the docs 
is absent from the code.

\textbf{Unused Parameter or Variable (3\%):} 
Unlike \textit{parameter mismatch}, the parameters specified in the documentation do match with those defined in the function signature. However, some parameters are not used in the code body at all; or, some C++ codebases use \texttt{GGML\_UNUSED(variable)} to avoid compiler warnings. 

The most prominent source of error varies by language. 
was \emph{Unimplemented Feature} the most prevalent in Java,
\emph{Functionality Mistmatch} was most common for Python, TypeScript and C++.

\begin{tcolorbox}[takeaway]
\textbf{RQ3 Takeaway.} 
11\% of developer-written code-documentation pairs in the multi-language dataset exhibit incorrectness inconsistencies. This provides a conservative lower-bound on the prevalence of developer-written code-documentation inconsistencies in real-world projects.
\end{tcolorbox}

\subsection{RQ4: Ablation Studies}  
\label{sec:rq4}

   

In this section, we evaluate the impact of different prompt variants on \toolname{}'s performance. 
Table~\ref{tab:ablations} shows the six variants of \toolname{}; they use different sets of building blocks, each defined below. The goal is to understand the impacts of two main design decisions in \toolname{}: \textit{local categorization} (LC) and \textit{external filtering} (EF). To understand LC, we look at the impact of chain-of-thought, as LC is akin to a specific form of chain-of-thought. To understand EF, we contrast with a more typical prompt engineering strategy of instructed filtering. 

\definecolor{darkgreen}{rgb}{0,0.5,0}
\definecolor{darkred}{rgb}{0.7,0,0}

\newcommand{\cmark}{\textcolor{darkgreen}{\ding{51}}} 
\newcommand{\xmark}{\textcolor{darkred}{\ding{55}}} 

\begin{table}[h!]
\centering
\small
\captionsetup{belowskip=1pt, aboveskip=5pt} 
\caption{\small Prompt and technique components in each studied variant. The \textsc{DC} variant is \toolname{}.}
\label{tab:ablations}
\scalebox{0.75}{
\setlength{\tabcolsep}{4pt}
\begin{tabular}{l c c c c c c c c c}
\toprule
\rotatebox{0}{\textbf{Variant}} & 
\rotatebox{0}{\textbf{Prefix}} & 
\rotatebox{0}{\textbf{Plain JSON}} & 
\rotatebox{0}{\textbf{CoT JSON}} & 
\rotatebox{0}{\textbf{Ins. Cat.}} & 
\rotatebox{0}{\textbf{Local Cat.}} &
\rotatebox{0}{\textbf{Ins. Filter}} & 
\rotatebox{0}{\textbf{Ext. Filter}} & 
\rotatebox{0}{\textbf{Type Label}} & 
\rotatebox{0}{\textbf{Summary}}  \\
\midrule
V1 & \cmark & \cmark &        &        &        &        &        &        &        \\
V2 & \cmark &        & \cmark &        &        &        &        &        &        \\
V3 & \cmark & \cmark &        & \cmark &        & \cmark &        &        &        \\
V4 & \cmark &        & \cmark & \cmark &        & \cmark &        &        &        \\
V6 & \cmark &        & \cmark & \cmark &        &        & \cmark & \cmark &          \\
V7 & \cmark &        & \cmark & \cmark &        &        & \cmark  & \cmark & \cmark \\
\textsc{DC} & \cmark  &       & \cmark &        & \cmark &        & \cmark &        & \cmark \\
\bottomrule
\end{tabular}}
\vspace{-10px}
\end{table}

\subsubsection{Variant Building Blocks}
\label{subsubsec:ablation-defs}
We consider nine building blocks for variant construction. 
\begin{enumerate}[leftmargin=*]
    \item \textbf{Prefix:}  As explained in Section ~\ref{subsubcsec: prefix}, \textit{Prefix} sets the model character, specifies the task, and indicates the input format. It is prefixed in all prompt variants.
    \item \textbf{Plain JSON:} Unlike \textit{CoT JSON}, \textit{Plain JSON} does not require the LLM to perform any chain-of-thought reasoning while generating outputs. It just needs to fill in the following key.
\begin{lstlisting}[language=text,breaklines=true, breakatwhitespace=false, xleftmargin=4pt, xrightmargin=4pt]
{"identified_inconsistency": "..."}

\end{lstlisting}  

    \item \textbf{CoT JSON:} 
    For each  inconsistency, the LLM must do chain-of-thought~\cite{wei2022chain} by: identifying which documentation/code snippets contain conflicting information, and provide an explanation. 

\begin{lstlisting}[language=text,breaklines=true, breakatwhitespace=false, xleftmargin=4pt, xrightmargin=4pt]
{"original_documentation_snippet_that_has_conflicting information_with_code": "...",
"original_code_snippet_that_has_conflicting_information_with_documentation": "...",  "explanation": "..."}
\end{lstlisting}   

    \item \textbf{Instructed Categorization:} 
    The prompt defines our three categories of inconsistencies (\S~\ref{sec:motivation}).
    \begin{lstlisting}[language=text,breaklines=true, breakatwhitespace=false, xleftmargin=4pt, xrightmargin=4pt]
You should focus on identifying three types of inconsistencies.
1. Over-promise: Some core parts of the documentation are not implemented in the code.
2. Direct mismatch: The code does not correctly implement what is mentioned in the documentation.
3. Under-promise: Some code is not documented or mentioned in the documentation.
\end{lstlisting}
    \item \textbf{Local Categorization:} Our JSON-based Local Categorization approach (Subtasks 3-5 in \S~\ref{subsubsec:system-prompt-json-schema}).

    \item \textbf{Instructed Filter:} 
    The prompt explicitly instructs the LLM not to report \textit{under-promises}.
\begin{lstlisting}[language=text,breaklines=true, breakatwhitespace=false, xleftmargin=4pt, xrightmargin=4pt]
Do not report inconsistencies in the third category, under-promise, in the final output.
\end{lstlisting}

    \item \textbf{External Filter:} 
    An external filtering module takes in the LLM output and filters out \textit{under-promises}  during the post-processing stage.

    \item \textbf{Type Label:} The LLM is must explicitly label the type of each inconsistency as either \textit{over-promise}, \textit{direct mismatch}, or \textit{under-promise}.
    
    
    \begin{lstlisting}[language=text,breaklines=true, breakatwhitespace=false, xleftmargin=4pt, xrightmargin=4pt]
{"inconsistency_type": "...", <other keys required for CoT JSON>}
\end{lstlisting}  
    \item \textbf{Code/Doc Summary:} The LLM is required to fill in two keys, \texttt{Documentation\_Summary} and \texttt{Code\_Summary}, before identifying and listing inconsistencies. 

\end{enumerate}

\begin{table}[h!]
\centering
\small
{\renewcommand{\arraystretch}{1}
\captionsetup{belowskip=1pt, aboveskip=5pt} 
\caption{\small \toolname{} vs. ablations on ablation dataset. \textit{F. Precision} refers to \textit{Function-level Precision}. \textit{I. Precision} denotes \textit{Inconsistency-level Precision}.
\textit{U.P. rate} means \textit{under-promise rate}; the lower, the better. Definitions of evaluation metrics can be found in Section~\ref{subsec:evaluation-metrics}.} 
\label{tab:python-dataset-ablation-result}
\scalebox{0.8}{
\setlength{\tabcolsep}{15pt}
\begin{tabular}{@{\extracolsep{3pt}} l c c c c c c c@{}}
  \toprule 
   & \textbf{V1} & \textbf{V2}
   & \textbf{V3} & \textbf{V4} & \textbf{V6} & \textbf{V7} & \textbf{\toolname{}}
   \\
  \midrule
   Flag rate & 94\% & 94\%  & 96\% & 98\% & 48\% & 39\%  & \cellcolor{gray!20}{14\%}\\
   F. Precision & 0.12 & 0.12 & 0.1 & 0.12 & 0.2 & 0.1 & \cellcolor{gray!20}{0.71}\\ 
   Recall & \cellcolor{gray!20}{1} & \cellcolor{gray!20}{1} & \cellcolor{gray!20}{1} & \cellcolor{gray!20}{1} & 0.83 & 0.33 & 0.71\\
   Accuracy & 0.17 & 0.17 & 0.14 & 0.14 & 0.6 & 0.58 & \cellcolor{gray!20}{0.94}\\
   F1 score & 0.22 & 0.22 & 0.18 & 0.21 & 0.32 & 0.15 & \cellcolor{gray!20}{0.77}\\
   U.P. rate  & 72\% & 44\% & 48\% & 28\% & 27\% & 50\% & \cellcolor{gray!20}{0\%}\\
   I. Precision & 0.07 & 0.14 & 0.07 & 0.1 & 0.17 & 0.09  & \cellcolor{gray!20}{0.7}\\
\bottomrule
\end{tabular}}}
\vspace{-8pt}
\end{table}

\begin{table}[t]
\centering
\rev{
\small
{\renewcommand{\arraystretch}{1}%
\captionsetup{belowskip=1pt, aboveskip=5pt}
\caption{\small \rev{Statistical evaluation of \toolname{}'s improvements over ablation variants using McNemar's test and Cohen's $g$, based on the results in Table~\ref{tab:python-dataset-ablation-result}. Cohen's $g$ ranges from 0 to 0.5; values of 0.25 or higher represent \textit{large} effect sizes.
}
}
\label{tab:ablation-dataset-stats}
\scalebox{0.8}{
\begin{tabular}{@{\extracolsep{1pt}} l c c c c @{}}
  \toprule
  \multicolumn{1}{c}{} & 
  \multicolumn{2}{c}{\textbf{\toolname{} vs. V4}} & 
  \multicolumn{2}{c}{\textbf{\toolname{} vs. V7}} \\
  \cmidrule(lr){2-3} \cmidrule(lr){4-5}
   & \textbf{p-value} & \textbf{Cohen's $g$} & \textbf{p-value} & \textbf{Cohen's $g$}  \\
  \midrule
   Flag rate & $1.1 \times 10^{-13}$ & 0.5 & $1.1 \times 10^{-2}$ & 0.28\\
   Accuracy   & $1.1 \times 10^{-10}$ & 0.45 & $6.6 \times 10^{-5}$ & 0.4\\
  \bottomrule
\end{tabular}}}
}
\vspace{-8pt}
\end{table}

\begin{table}[t!]
\begin{minipage}[t]{0.51\textwidth}
\centering
\small
{\renewcommand{\arraystretch}{1}
\captionsetup{belowskip=1pt, aboveskip=5pt} 
\caption{\small \toolname{} vs. ablations: flag rates on the main Python, TypeScript, C++, and Java datasets across ablation variants. \vspace{1.3em}}
\label{tab:wild-dataset-ablation-result}
\scalebox{0.85}{
\setlength{\tabcolsep}{5pt}
\begin{tabular}{@{\extracolsep{1pt}}  l c c c c c c c@{}}
  \toprule 
   & \textbf{V1} & \textbf{V2}
   & \textbf{V3} & \textbf{V4} & \textbf{V6} & \textbf{V7} & \textbf{\toolname{}}
   \\
  \midrule
   Python & 95\% & 95\% & 97\% & 97\% & 95\% & 56\% &  \cellcolor{gray!20}19\%  \\
   TypeScript  & 93\% & 90\% & 93\% & 94\% & 58\% & 52\% & \cellcolor{gray!20}{16\%}\\
   C++  & 90\% & 90\% & 93\%  & 92\% & 62\% & 60\%& \cellcolor{gray!20}{14\%}\\
   Java  & 94\% & 92\% & 91\% & 91\% & 67\% & 57\% & \cellcolor{gray!20}{18\%}\\
\bottomrule
\end{tabular}}}

\end{minipage}
\hfill
\begin{minipage}[t]{0.45\textwidth}
{\renewcommand{\arraystretch}{1}
\captionsetup{belowskip=1pt, aboveskip=5pt} 
\caption{\small 
\toolname{} vs. ablations: flag rates across ablation variants with GPT4.1. ($\pm X\%$) refers to the difference between the adjacent value and the corresponding one in Table~\ref{tab:wild-dataset-ablation-result}.  
} 
\label{tab:ablation-result-gpt}
\scalebox{0.8}{
\begin{tabular}{@{\extracolsep{5pt}} l c c c @{}}
  \toprule 
    & \textbf{V4} & \textbf{V7} & \textbf{\toolname{}}
   \\
  \midrule
   Python  & 91\%(-6\%) & 60\%(+4\%) & \cellcolor{gray!20}{24\%(+5\%)} \\
   TypeScript  & 82\% (-12\%) & 47\% (-5\%) & \cellcolor{gray!20}{17\% (+1\%)} \\
   C++  & 90\% (-2\%) & 51\% (-9\%) & \cellcolor{gray!20}{20\% (+6\%)}  \\
   Java  & 86\% (-5\%) & 60\% (+3\%) & \cellcolor{gray!20}{24\% (+6\%)}  \\
   
\bottomrule
\end{tabular}}}
\end{minipage}
\vspace{-2em}
\end{table}


\begin{table}[t!]
\begin{minipage}[t]{0.49\textwidth}
\centering
\rev{
\small
{\renewcommand{\arraystretch}{1.15}%
\captionsetup{belowskip=15pt, aboveskip=5pt}
    \caption{\small \rev{Statistical evaluation of \toolname{}'s improvements over ablation variants using McNemar's test and Cohen's $g$, based on the results in Table~\ref{tab:wild-dataset-ablation-result}. Cohen's $g$ ranges from 0 to 0.5; values of 0.25 or higher are typically considered large effects.}}
\label{tab:llama-flag-rates-stats}
\scalebox{0.8}{
\begin{tabular}{@{\extracolsep{0pt}}lcccc @{}}
  \toprule
   & 
  \multicolumn{2}{c}{\textbf{\toolname{} vs V4}} & 
  \multicolumn{2}{c}{\textbf{\toolname{} vs V7}} \\
  \cmidrule(lr){2-3} \cmidrule(lr){4-5}
   & \textbf{p-value} & \textbf{Cohen's $g$} 
   & \textbf{p-value} & \textbf{Cohen's $g$} \\
  \midrule
  Python   & $2.0 \times 10^{-118}$ &  0.5 & $1.2 \times 10^{-43}$ & 0.4  \\
  TypeScript   & $4.9 \times 10^{-114}$  & 0.5  & $9.7 \times 10^{-43}$ & 0.4  \\
  C++  & $2.5 \times 10^{-116}$ & 0.5 & $5.1 \times 10^{-54}$ & 0.4 \\
  Java & $2.4 \times 10^{-108}$  & 0.5  & $3.2 \times 10^{-43}$ & 0.4  \\
  \bottomrule
\end{tabular}}}
}

\end{minipage}
\hfill
\begin{minipage}[t]{0.49\textwidth}
\rev{
\small
{\renewcommand{\arraystretch}{1.15}%
\captionsetup{belowskip=15pt, aboveskip=5pt}
\caption{\small \rev{Statistical evaluation of \toolname{}'s improvements over ablation variants using McNemar's test and Cohen's $g$, based on the results in Table~\ref{tab:ablation-result-gpt}. Cohen's $g$ ranges from 0 to 0.5; values of 0.25 or higher are typically considered large effects.} }
\label{tab:gpt-flag-rates-stats}
\scalebox{0.8}{
\begin{tabular}{@{\extracolsep{0pt}}lcccc @{}}
  \toprule
   & 
  \multicolumn{2}{c}{\textbf{\toolname{} vs V4}} & 
  \multicolumn{2}{c}{\textbf{\toolname{} vs V7}} \\
  \cmidrule(lr){2-3} \cmidrule(lr){4-5}
   & \textbf{p-value} & \textbf{Cohen's $g$} 
   & \textbf{p-value} & \textbf{Cohen's $g$} \\
  \midrule
  Python   & $5.7 \times 10^{-101}$ & 0.5 & $2.1 \times 10^{-47}$ & 0.47  \\
  TypeScript   & $7.5 \times 10^{-96}$  & 0.5 & $8.4 \times 10^{-43}$ & 0.49  \\
  C++  & $8.5 \times 10^{-109}$ & 0.5  & $2.1 \times 10^{-50}$ & 0.5 \\
  Java & $1.5 \times 10^{-92}$  & 0.5  & $4.2 \times 10^{-50}$ & 0.49 \\
  \bottomrule
\end{tabular}}}
}
\end{minipage}
\vspace{-1em}
\end{table}

\subsubsection{Results} 
Table~\ref{tab:python-dataset-ablation-result} presents the ablation study results on the 
ablation dataset across seven evaluation metrics. 
Table~\ref{tab:wild-dataset-ablation-result} shows the flag rates on the Python, TypeScript, C++, and Java datasets. 
\rev{Table~\ref{tab:ablation-dataset-stats} and Table~\ref{tab:llama-flag-rates-stats} report McNemar's test~\cite{mcnemar1947note} results and Cohen's $g$~\cite{cohen2013statistical} values to evaluate \toolname{}'s improvements over the ablation variants in terms of accuracy and flag rate. All p-values are well below 0.01. Cohen's $g$ values range from 0.28 to 0.50, exceeding the conventional threshold of 0.25 for a large effect. This indicates that \toolname{}'s improvements over the ablation variants in flag rate and accuracy are both  statistically significant and practically substantial.}

Compared to all six variants, \toolname{} achieves up to a 5.1$\times$ improvement in F1 score, 6.7$\times$ in accuracy, 7.1$\times$ in function-level precision, and 10$\times$ in inconsistency-level precision on the ablation dataset. 
It also reduces the \textit{under-promise} rate from 72\% (V1) to 0\% and the flag rate from 98\% (V4) to 14\%. Similarly, the average flag rate 
on  Python, TypeScript, C++, and Java  
 goes
 from 94\% to 17\%.

\textbf{Effects of Chain-of-Thought (V1 vs. V2 / V3 vs. V4).} 
\textit{CoT JSON} 
guides the LLM use chain-of-thought reasoning while generating outputs.
V2 adds \textit{CoT JSON} to V1, and V4 adds \textit{CoT JSON} to V3. 
Adding \textit{CoT JSON} reduces under-promise rates in both cases: from 72\% (V1) to 44\% (V2), and from 48\% (V3) to 28\% (V4). We also see consistent improvements in inconsistency-level precision, going up from 0.07 (V1) to 0.14 (V2), and from 0.07 (V3) to 0.10 (V4).



\textbf{Effects of Instructed Categorization and Instructed Filter (V1 vs. V3 / V2 vs. V4).} 
V3 and V4  
add \textit{Instructed Categorization} and \textit{Instructed Filter} to V1 and V2, respectively. This enhances the prompt with the three inconsistency categories and instructions to filter out \textit{under-promises}. 
We do see the under-promise rate reduce: from 72\% (V1) to 48\% (V3) and from 44\% (V2) to 28\% (V4). However, the LLM still struggles to mitigate the under-promise rate:  
 48\% and  28\% is still quite high.



\textbf{Effects of External Filter (V6 vs. V4).}
Instead of telling the LLM not to report \textit{under-promises} through the prompt (V4), V6 prompts the LLM to label each detected inconsistency with its category (i.e., \textit{over-promise}, \textit{direct mismatch}, or \textit{under-promise}). It then filters out inconsistencies labelled with \textit{under-promise} 
during the post-processing stage. V6 reduces the average flag rate from 92\% to 62\% on the TypeScript, C++, and Java datasets. It improves six out of seven evaluation metrics on the ablation dataset. Notably: reducing the flag rate from 98\% (V4) to 48\% (V6) and increasing accuracy from 0.14 to 0.6---due to a substantial increase in true negatives. But, there is no significant decrease in the under-promise rate. This is partly because V6 mislabels some \textit{under-promises} as \textit{direct mismatches}, 
so that some real \textit{under-promises} cannot be filtered out. In contrast, \toolname{}, with the \textit{Local Categorization}, substantially reduces the instances of mislabelling. 

\textbf{Effects of Local Categorization (\toolname{} vs. V7)} 
V7 adds documentation and function summaries to V6. Like V6 and V7, \toolname{} employs an \textit{external filter} to drop under-promises during the post-processing stage. However, unlike V6/V7, \toolname{} adopts \textit{local categorization} (LC), where the LLM is not required to ``remember'' the definitions of the inconsistency categories. 
Instead, it must answer the categorization questions for each check-in key in the predefined JSON schema. \toolname{}
 outperforms V7 on all evaluation metrics, and V6 on all metrics except recall. It reduces the flag rate from 39\% to 14\% and the under-promise rate from 50\% to 0\%, and increases function-level precision from 0.1 to 0.71, accuracy from 0.58 to 0.94, F1 score from 0.15 to 0.77, and inconsistency-level precision from 0.09 to 0.70. This shows the effectiveness of \textit{local categorization}.

\begin{tcolorbox}[takeaway]
\textbf{RQ4 Takeaway.} 
\emph{Local Categorization, External Filtering} (LCEF) is highly effective at reducing \toolname{}'s under-promise and flag rates. 
Both LC and EF greatly decrease flag rates, but LC is key to decreasing under-promise rate and increasing precision. 

\end{tcolorbox}



\subsection{RQ5: Generalization to Other LLMs}  
\label{sec:rq5}

The previous research question (ref. Section~\ref{sec:rq4}) shows that the \emph{Local Categorization, External Filtering} (LCEF) approach using LLaMA3.1-70B is highly effective at reducing flag rates.
This research question investigates how our approach generalizes to other LLMs. 

We select two representative ablation variants introduced in Section~\ref{sec:rq4}: V4 and V7. V4 instructs LLMs to filter out \textit{under-promise} inconsistencies in the prompt (i.e., \textit{Instructed Filter}). V7 replaces the \textit{Instructed Filter} with the \textit{Type Label} and \textit{External Filter} components, i.e., asks the LLM to explicitly categorize each inconsistency. 
We run these variants and \toolname{} using GPT4.1 on our datasets. 

Table~\ref{tab:ablation-result-gpt} shows the results. We see the same trend of decreased flag rate, from V4 to V7 to \toolname{}, 
observed with LLaMA3.1-70B in RQ4 (ref. Section~\ref{sec:rq4}). 
\rev{We use McNemar's test~\cite{mcnemar1947note} and Cohen's $g$~\cite{cohen2013statistical} to evaluate \toolname{}'s flag-rate improvements over the ablation variants. As shown in Table~\ref{tab:gpt-flag-rates-stats}, all p-values are near zero, and Cohen's $g$ ranges from 0.47 to 0.50, exceeding Cohen's $g$ threshold for large effect sizes (0.25). Thus, the flag-rate improvements achieved by \toolname{}-GPT4.1 are both statistically significant and practically substantial. }

Comparing Table~\ref{tab:wild-dataset-ablation-result} with Table~\ref{tab:ablation-result-gpt}, we see that \toolname{}-GPT4.1  has slightly higher flag rates than \toolname{}-LLaMA3.1. 
But V4-GPT4.1 has lower flag rates than V4-LLaMA3.1. This suggests the standard ``define categories and instruct to filter''  approach is more effective on GPT4.1. Overall, the results show that a newer-generation model still suffers from the high flag rate we observed with LLaMA3.1, and that LCEF is still effective at reducing it, regardless of the model used.



\begin{tcolorbox}[takeaway]
\textbf{RQ5 Takeaway.} 
 Even a proprietary model (GPT4.1) has unacceptably high flag rates when using standard prompting techniques. \toolname{} is effective at reducing high flag rates with this proprietary model, suggesting it is not over-fit to one specific model.
\end{tcolorbox}

\section{Discussion}
\label{sec:discussion}


%
Our evaluation shows that code-documentation inconsistency errors are not uncommon in real-world developer-written code (RQ3).
Our ablation study (RQ4) demonstrates how LCEF dramatically decreases flag- and under-promise rates, but also greatly improves accuracy to over 90\%. 
We think  low flag and under-promise rates are important for developers wanting to use a tool detecting code-documentation inconsistencies
---otherwise they  will be overwhelmed with unimportant warnings.

\toolname{} performance contrasts starkly with our evaluation of SOTA (RQ1). 
C4RLLaMA  achieved 94\% precision and 90\% accuracy on a widely-used, but synthetic, dataset. But, it flagged every function in our dataset---made of  code-documentation pairs that co-occur in real codebases---as inconsistent with its documentation\rev{, leading to a precision of only 0.05-0.14}. 
This result aligns with our ablation study, where our first uses of LLMs to find inconsistencies had flag rates over 90\%. 
However, it shows a danger with relying on automatically-labelled datasets and binary labels in building software engineering tools. 

\revdel{A shortcoming of our comparison to the SOTA is that, due to the high volume of C4RLLaMA-reported inconsistencies, we could not manually analyze the correctness of  the results.
However, the C4RLLaMA paper reported that 65\% of rectifications were correct.
Using their reported 65\% correctness rate and applying it to our data, C4RLLaMA could be estimated to have 405 false flags, over 4$\times$ the 
84 false flags \toolname{} surfaced. }


\textbf{Ranking results.}
Given the surprisingly large numbers of inconsistencies found in real-world code (RQ3), flagging this many results for a developer could still be overwhelming.
We could reduce this by ranking inconsistencies, to help developers focus on only those that are most impactful.
This would help them to maximize any effort invested into improving the documentation consistency within their project.
Such improvements could help both future developers reading the documentation, as well as any emerging LLM-based development tools that rely on accurate documentation as the basis for constructing prompts for generating implementations.

\textbf{Overlap between different inconsistencies.} 
When examining our results, we noted that \toolname{} sometimes outputs the same inconsistency into two categories. 
This happens most often with an inconsistency being listed as both an over-promise and direct mismatch, though sometimes an under-promise was also listed as direct mismatch (e.g., Fig~\ref{fig:underpromise-example2}). 
This suggests that Local Categorization's effectiveness depends on how well the Check-In Key question describes the target category. Upon analysis, our Check-In Key for direct mismatches, \textit{``Does the code correctly implement what is mentioned  in the documentation?''}, is semantically vague: ``correctly implement'' can also capture the notion of over-promise. This suggests that the success of LCEF may depend on whether a brief, unambiguous check-in key can be constructed.
\section{Conclusion}
\label{sec:conclusion}

In this paper, we have introduced \toolname{}, which flags and explains incorrectness inconsistencies between functions and their documentation.
Examining developer-written source code and their documentation from 20 real-world repositories across four programming languages,
\toolname{} flags a low number of inconsistencies ($< 18\%$), meaning developers do not need to examine many results from this analysis.
This is much lower than the over 90\% flag rates of standard prompting techniques, or 100\% flag rates of state-of-the-art. 
Our approach demonstrates that detecting inconsistency errors between real-world developer-written function-level documentation and their source code is possible. 
We believe tools such as \toolname{} will enable developers to better trust the documentation in their systems and once again rely on documentation as an accurate and useful form of abstraction crucial for building and evolving their systems.


\section*{Acknowledgments}
Thanks to the anonymous reviewers whose suggestions have improved the paper. We acknowledge the support of the Natural Sciences and Engineering Research Council of Canada (NSERC), [CRC-2024-00299, RGPIN-2023-04478]. 

\section{Data Availability}
\label{sec:data_availability}
The artifact is available at
\url{https://github.com/SnowPhoebe/DocPrism}
and
\url{https://doi.org/10.5281/zenodo.21449177}.
It includes the datasets containing all of the functions and documentation we examined, the prompts (including ablation prompts), \toolname{} outputs for the tested functions, baseline outputs, our manually-determined labels, and details of the 22 projects from which the datasets were collected. It also contains representative examples for each true positive or false positive category, as identified by our tool \toolname{}. Examples are identified as codebase-n, corresponding to our n\textsuperscript{th} function in that codebase's dataset.


\balance
\bibliographystyle{ACM-Reference-Format}
\bibliography{citations}


\end{document}